\documentclass[%
reprint,
superscriptaddress,
amsmath,amssymb,
aps,
prl,
]{revtex4-2}
\usepackage{graphicx}% Include figure files
\usepackage{bm}% bold math
\usepackage{amsmath}
\usepackage{xcolor}
\newcommand{\RNum}[1]{\uppercase\expandafter{\romannumeral #1\relax}}
\begin{document}

\title{Why and when merging surface nanobubbles jump}
\author{Yixin Zhang}
%\email{y.zhang-11@utwente.nl}
%\email{Y.Zhang.-11@utwente.nl}
  \affiliation{Physics of Fluids Group, Max Planck Center Twente for Complex Fluid Dynamics and J. M. Burgers Centre for Fluid Dynamics, University of Twente, P.O. Box 217, 7500 AE Enschede, The Netherlands}
\author{Xiangyu Zhang}
  \affiliation{Department of Architecture and Civil Engineering, City University of Hong Kong, Kowloon, Hong Kong, PR China}
\author{Detlef Lohse}
\email{d.lohse@utwente.nl}
  \affiliation{Physics of Fluids Group, Max Planck Center Twente for Complex Fluid Dynamics and J. M. Burgers Centre for Fluid Dynamics, University of Twente, P.O. Box 217, 7500 AE Enschede, The Netherlands}
  \affiliation{Max Planck Institute for Dynamics and Self-Organization, 37077 Göttingen, Germany}
\begin{abstract}
Gas bubble accumulation on substrates reduces the efficiency of many physicochemical processes, such as water electrolysis. For microbubbles, where buoyancy is negligible, coalescence-induced jumping driven by the release of surface energy provides an efficient pathway for their early detachment. At the nanoscale, however, gas compressibility breaks volume conservation during coalescence, suppressing surface energy release and seemingly disabling this detachment route. Using molecular dynamics simulations, continuum numerical simulations, and theoretical analysis, we show that surface nanobubbles with sufficiently large contact angles can nevertheless detach after coalescence. In this regime, detachment is powered by the release of pressure energy associated with nanobubble volume expansion. This finding thus establishes a unified driving mechanism for coalescence-induced bubble detachment across all length scales.
\end{abstract} 
\maketitle 
 \newpage
Bubbles are present in many physicochemical and engineering processes\,\citep{lohse2018bubble}, such as water electrolysis\,\citep{PhysRevLett.123.214503,zhang2024threshold,angulo2020influence,bashkatov2024performance,zhao2019gas,shih2022water}, boiling heat transfer\,\citep{li2021liquid,gallo2023nanoscale,park2024coalescence,prosperetti2017vapor}, micro- and nanofluidics\,\citep{kavokine2021fluids}, electrochemical reduction of carbon dioxide\,\citep{singh2015effects}, and membrane-based separation. These bubbles form on interfaces and hinder the transfer of mass, momentum, or heat, thereby reducing efficiency. For instance, bubbles nucleated on electrode surfaces can block electrolytes' access to electrodes\,\citep{shih2022water,PhysRevLett.123.214503,zhang2024threshold,angulo2020influence,bashkatov2024performance,zhao2019gas}, which decreases reaction area and increases overpotential. Large bubbles can often detach under buoyancy, but smaller ones with negligible buoyancy remain strongly attached. Many strategies have been proposed to promote detachment of small bubbles, including shear flow, acoustic forcing, and coalescence\,\citep{he2023strategies}. Among these, coalescence-induced detachment\,\citep{soto2018coalescence,lv2021self,li2025size,demirkir2025jump,lu2025combined,zhao2022coalescence,iwata2022coalescing}, being passive and self-propelled, has attracted particular attention for mitigating bubble blockage and improving process performance.

Coalescence-induced jumping was first observed for droplets on superhydrophobic substrates\,\citep{boreyko2009self}; see\,\citep{boreyko2024jumping} for a comprehensive review. Upon merging, the reduced surface area releases energy, while the total droplet volume remains constant. The substrate breaks the symmetry of coalescence, allowing part of this energy to convert into kinetic energy and drive detachment. For large droplets the jumping velocity follows the capillary–inertial scaling $u_{ci}\propto \sqrt{\gamma/(\rho_l R)}$\,\citep{PhysRevLett.95.164503,lv2013condensation,enright2014coalescing,liu2014numerical,mouterde2017merging,vo2019critical,bird2021coalescence,eggers2024coalescence}, with a prefactor of about 0.2, where $R$ is the radius, $\gamma$ is the surface tension, and $\rho_l$ is the liquid density. At smaller scales, however, viscous dissipation dominates and the scaling fails\,\citep{liu2014numerical}, as quantified by the increasing Ohnesorge number $Oh=\mu/\sqrt{\gamma\rho_l R}$, i.e., the nondimensionalized dynamic viscosity $\mu$.
\begin{figure}[t]
\includegraphics [width=\linewidth]{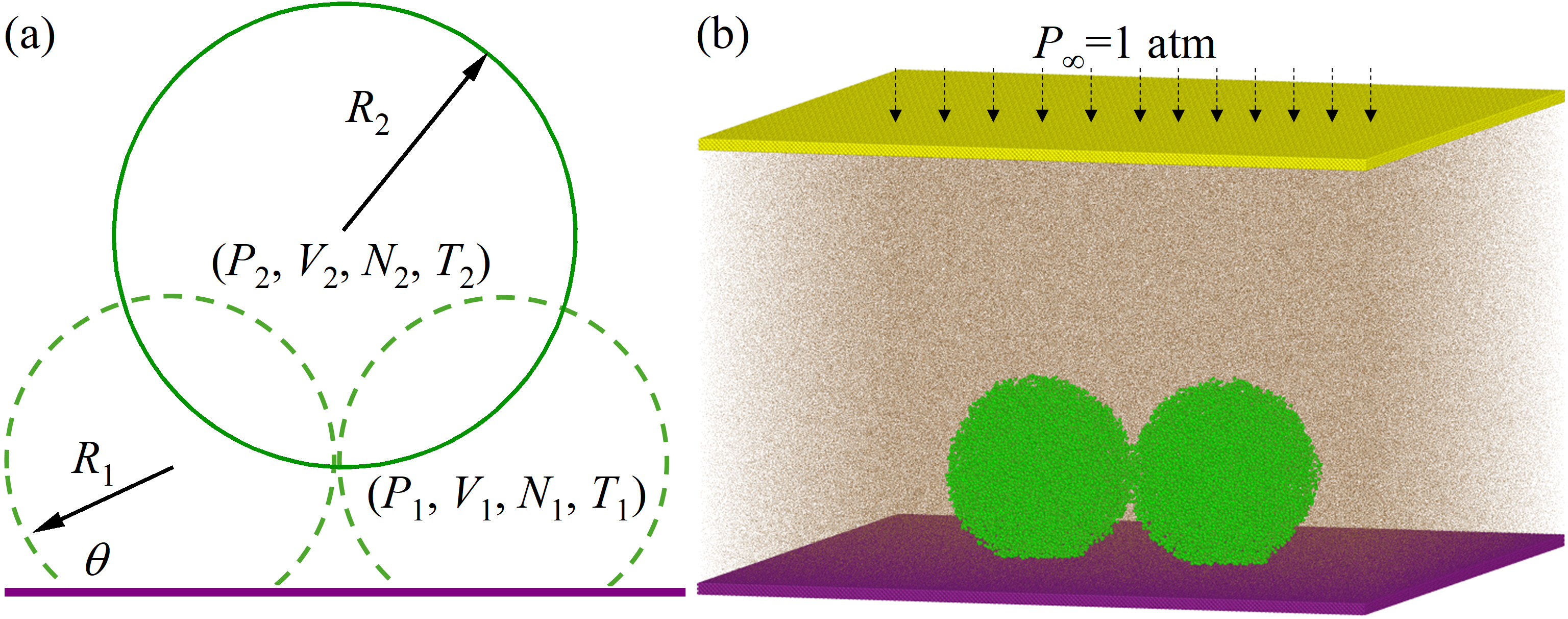}
\caption{\label{fig1} (a) Sketch of the coalescence-induced nanobubble detachment. The nanobubbles (dashed cap) before coalescence are identical for simplicity and have a gas-side contact angle $\theta$ and a radius of curvature $R_1$, and the thermodynamic state $(P_1,V_1, N_1,T_1)$. The merged nanobubble (solid sphere) after coalescence has a radius of $R_2$ and the thermodynamic state $(P_2,V_2,N_2,T_2)$. (b) A perspective view of the initial configuration of nanobubble coalescence on the hydrophilic substrate in MD simulations. The system's condition is maintained at $T=300$ K and $P_{\infty}=1$ atm.}
\end{figure}

This mechanism has also been extended to microbubble (MBs) detachment, particularly for electrolytic gas bubbles on hydrophilic electrodes\,\citep{lv2021self,li2025size,demirkir2025jump,lu2025combined}.
As gas bubbles are often generated by nanoscale nucleation under supersaturated environments\,\citep{lohse2015surface,chen2014electrochemical}, seeking even earlier detachment right after their nucleation is of course crucial. At the nanoscale, however, whether coalescence can trigger detachment remains unclear. Nanobubbles (NBs) experience strong Laplace pressure, leading to significant gas compressibility that invalidates volume conservation, the basis for surface energy release. Combined with the high viscous dissipation, these effects raise fundamental doubts on whether nanobubble detachment by coalescence would be possible.
\begin{figure*}[t]
\includegraphics [width=\linewidth]{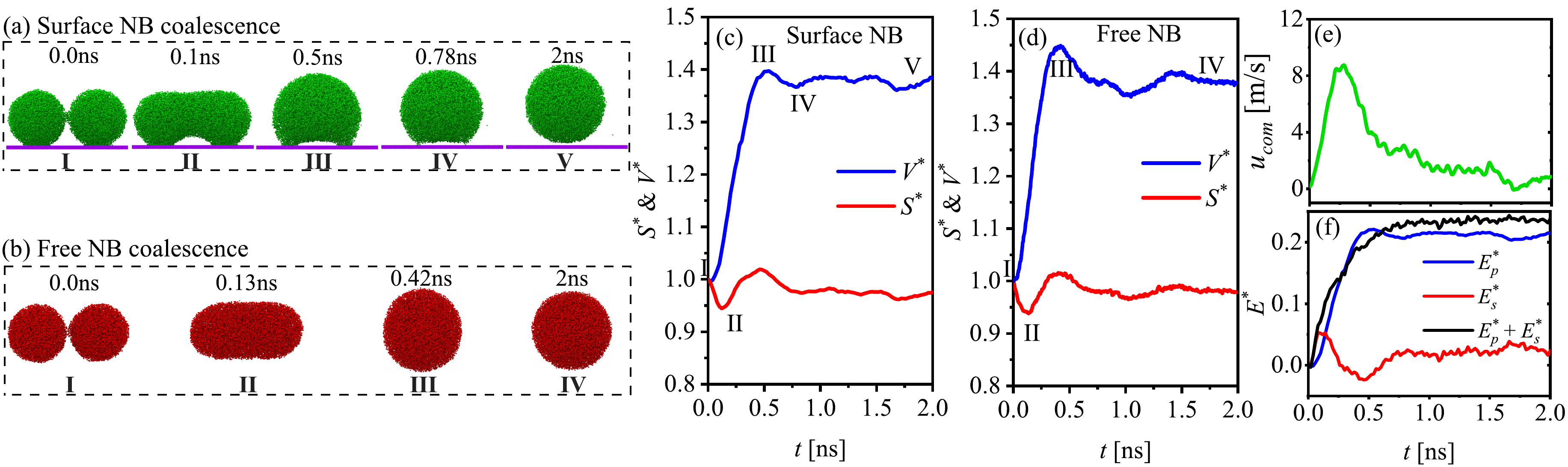}
\caption{\label{fig2}(a) Coalescence process of surface nanobubbles (NBs) in MD simulations, characterized by five typical states: I, initial contact; II, top liquid bridge flattening after expansion; III, maximum NB volume; IV, NB departure from the substrate; V, equilibrated NB.  
(b) Coalescence process of free NBs in MD simulations, characterized by four typical states (no departure): I, initial contact; II, liquid bridges flattening after expansion; III, maximum NB volume; IV, equilibrated NB. Notably, the flattening of liquid bridges corresponds to the minimum of surface area. (c) Surface NB's volume and area changes, normalized by their initial values. (d) Free NB's normalized volume and area changes. (e) Instant surface NB's center-of-mass velocity $u_{com}$ measured from MD. (f) Energy release (nondimensional) during surface NBs' coalescence in MD. $E_p$ is the pressure energy and $E_s$ is the surface energy. 
}
\end{figure*}

This Letter addresses this very question, considering the concerns outlined above. Using molecular dynamics simulations, continuum numerical simulations, and mathematical modeling, we find that gas compressibility greatly reduces the release of surface energy at the nanoscale. Nevertheless, nanobubbles with sufficiently large contact angles can still overcome viscous dissipation and detach, but now driven by the release of pressure energy as the merged nanobubbles expand to a new equilibrium state. By combining the role of surface energy, which dominates at the microscale and above, with pressure energy, which dominates at the nanoscale, this work thus establishes a unified driving mechanism for coalescence-induced bubble detachment across all length scales.

\emph{MD Simulations}.\,--- Molecular dynamics (MD) simulations are employed as virtual experiments to study the coalescence of nanobubbles. The simulations are performed using the open source code LAMMPS\,\cite{plimpton1995fast}. Details of parameters and strategies are provided in the Supplemental Materials (SM-I). As shown in Fig.\,\ref{fig1}(b), the molecular system consists of water molecules (orange), gas atoms (green), substrate atoms (purple), and atoms of a ``piston'' plate (yellow). The piston is used to maintain the far-field pressure at 1 atm, while the fluids' (liquid and gas) temperature is maintained at 300 K using the Nosé–Hoover thermostat. Notably, using a constant-energy simulation for the fluids with only the substrate thermostated yields almost the same results. Water is modeled using the mW potential\,\cite{molinero2009water}, which gives a surface tension $\gamma = 66$~mN/m and a viscosity $\mu=0.35$ mPa\,s\,\citep{zhang2025motion}. The gas, modeled with the standard 12-6 Lennard-Jones (LJ) potential, has a density $\rho_{\infty} = 1.15$~kg/m$^3$ at 1 atm and 300 K, representing nitrogen. The static water-solid contact angle is set to about $150^\circ$. Each of the two initial nanobubbles, which are assumed to be of equal size for simplicity, has a radius of 10~nm and contains 13356 gas atoms, within a system of roughly 10 million atoms. Though the radii of NBs should grow over time in the supersaturated environment during water splitting\,\citep{zhang2024threshold}, coalescence occurs much faster than mass diffusion. The latter is thus neglected by tuning the gas–water interaction to yield extremely low gas solubility, a common strategy in nanobubble cavitation modeling\,\citep{shekhar2013nanobubble}. For comparison, we also simulate the coalescence of two free NBs, i.e., far away from any interfaces, as their energy components are essentially the same as those of surface NBs.%It is noted that the static contact angle may be affected by various factors, including the Tolman length, line tension, and pressure dependent gas-solid interfacial energy. To minimize these effects, as well as Brownian motion,

Initially, the two NBs are separated by approximately $0.8\,\mathrm{nm}$. Figures~\ref{fig2}(a) [surface NBs] and \ref{fig2}(b) [free NBs] show consecutive MD snapshots of typical coalescence states after thermal fluctuations bring the NBs into contact, see the caption for the meaning of these states. The video of the surface NB simulation is also provided (Movie S1). Despite an estimated high $Oh=0.4$ where according to a continuum approach for MBs no detachment should be possible\,\citep{demirkir2025jump}, the merged surface NB in Fig.\,\ref{fig2}(a) does detach, accompanied by a rapid increase in its center-of-mass velocity $u_{com}$, as shown in Fig.\,\ref{fig2}(e). For free NBs in Fig.\,\ref{fig2}(b), symmetry leads to negligible movement of the center of mass. At first glance, the jumping of the surface NB might be attributed to surface energy release, as discussed in the Introduction for droplets and MBs. However, measurements of surface area change after NBs' coalescence show that it is minimal (red lines) while the volume (blue lines) increases by $\sim 40\%$ in Figs.\,\ref{fig2}(c) and \ref{fig2}(d). This is very different from MBs' coalescence, which would exhibit a $\sim 20.6\%$ area reduction upon merging and a constant bubble volume over time. These unusual NBs' coalescence behaviors will now be analyzed.
\begin{figure}[t]
\includegraphics [width=0.9\linewidth]{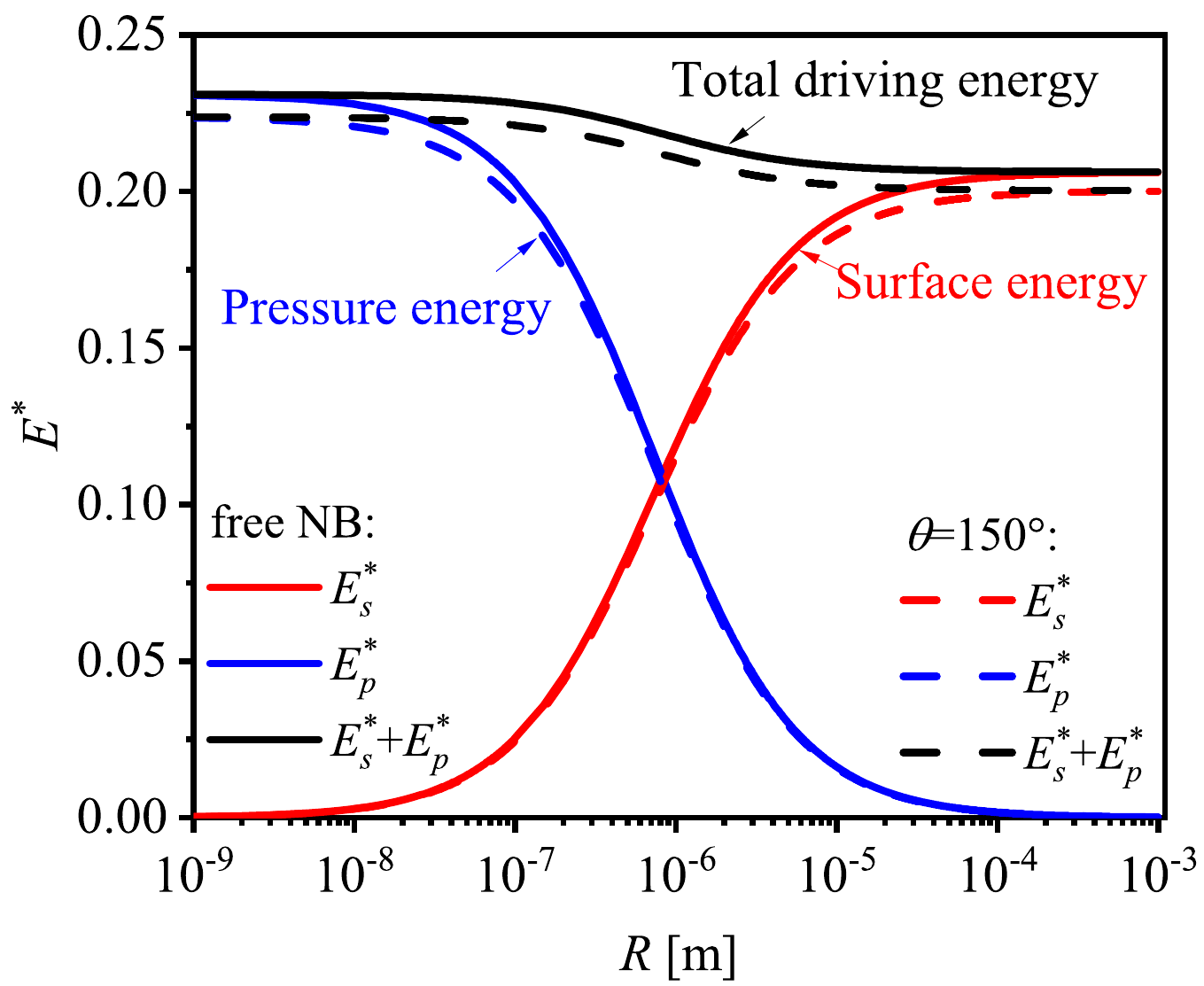}
\caption{\label{fig3} Energy release as a function of bubble radii, nondimensionalized by the total surface energy $8\pi\gamma R^2$. Solid lines are for free NBs or surface NBs with $\theta=180^{\circ}$. Dashed lines are for surface NBs with $\theta=150^{\circ}$.}
\end{figure}

\emph{Energy analysis for merging nanobubbles}.\,--- The two identical gas nanobubbles before coalescence and the new nanobubble after coalescence at equilibrium all follow the ideal gas law $P_1 V_1=N_1k_BT_1$ and $P_2 V_2=N_2k_BT_2$, where $P,V,N$ are the pressure, volume, and number of gas atoms of the bubble. $k_B$ is the Boltzmann constant. Assuming constant temperature ($T_1 = T_2$), mass conservation $2N_1 = N_2$ gives $2P_1 V_1 = P_2 V_2$. Ignoring the small vapor pressure, the mechanical equilibrium condition for equilibriated bubbles implies  
\begin{equation} \label{eq_pv}
    2\left(P_{\infty}+\frac{2\gamma}{R_1}\right)V_1=\left(P_{\infty}+\frac{2\gamma}{R_2}\right)V_2.
\end{equation}
Balancing the external pressure and the Laplace pressure defines a characteristic length scale $l=2\gamma /P_{\infty}\approx 1.4~\mathrm{\mu m}$, using $\gamma=72$ mN/m and $P_{\infty}=1$ atm. For MBs and bubbles whose sizes are well above $\ell$, $P_{\infty} \gg   2\gamma/R_1$, this leads to the usual volume conservation $2V_1 = V_2$. In the case of free bubbles or bubbles with $\theta = 180^{\circ}$, their geometry correspondingly gives $R_2 = 2^{1/3} R_1$. The nondimensional surface area reduction is then $\Delta S^* =\left( S_1-S_2\right)/\left(8\pi R_1^2\right) =\left(2-2^{2/3} \right)/2=20.6\%$, which drives the detachment of MBs after their coalescence\,\citep{lv2021self,li2025size,demirkir2025jump,lu2025combined}. However, for NBs with sizes well below $\ell$, such as those simulated by our MD simulations, $P_{\infty} \ll  2\gamma/R_1$, so that one has $2V_1/R_1=V_2/R_2$ instead, which means $R_2=2^{1/2}R_1$ and $\Delta S^* = 0$. This thus explains the negligible change of surface area after coalescence observed for free and surface NBs in Figs.\,\ref{fig2}(d) and \ref{fig2}(c), respectively. 

The above discussion of area change for free bubbles corresponds to two limiting cases derived from the ideal gas law. In Fig.\,\ref{fig3}, the red solid line shows the actual values obtained by solving Eq.\,\eqref{eq_pv} numerically. Indeed, as the bubble radius decreases from the millimeter scale to the nanometer scale, the nondimensional surface energy change (equivalent to the area change) decreases from $20.6\%$ to nearly zero.

We emphasize that the nearly zero reduction of surface area (energy) for free NBs' coalescence also applies to the merging of surface NBs with a contact angle $\theta$, see again Fig.\,\ref{fig2}(c). The interfacial energy for a surface bubble is $\gamma \left(A_{lg}-A_{sg}\cos \theta\right)$\,\citep{giacomello2012metastable,xiang2017ultimate}, based on Young's relation, which suggests an equivalent interfacial area $A_{lg}-A_{sg}\cos \theta$, where `$lg$' is short for `liquid-gas' and `$sg$' for `solid-gas'. Using the bubble's geometry, the surface energy release is thus
\begin{eqnarray}
     & E_s=\gamma \Delta S  \nonumber \\
   & =\gamma \left[4\pi R_1^2\left(1-\cos \theta\right)-2\pi R_1^2 \sin \theta^2 \cos \theta-4\pi R_2^2\right]\!.
\end{eqnarray}
%The area change for surface bubbles is $\Delta S= 4\pi R_1^2\left(1-\cos \theta\right)-2\pi R_1^2 \sin \theta^2 \cos \theta-4\pi R_2^2$.
For NBs' coalescence in the small radius limit of  $2V_1/R_1=V_2/R_2$, $R_2=R_1\sqrt{ \left(2+\cos\theta\right)\left(1-\cos \theta\right)^2/2}$ is obtained. Then the area change after surface NBs' coalescence is indeed found to be zero:
\begin{eqnarray}
&\Delta S_{NB}= 4\pi R_1^2\left[{1-\cos \theta} -\frac{1}{2}\sin \theta^2 \cos \theta\right.\nonumber\\
& -\left.\frac{\left(2+\cos\theta\right)\left(1-\cos \theta\right)^2}{2} \right] \equiv 0.
\end{eqnarray}
%The new volume is 
However, for MBs' coalescence in the large radius limit of $2V_1=V_2$, $R_2=R_1\left[\left(2+\cos\theta\right)\left(1-\cos \theta\right)^2/2\right]^{1/3}$. The usual area change after MBs' coalescence is
\begin{eqnarray}\label{eq_dsmb}
&\Delta S_{MB}= 4\pi R_1^2\left\{{1-\cos \theta} -\frac{1}{2}\sin \theta^2 \cos \theta\right.\nonumber\\
& -\left.\left[\frac{\left(2+\cos\theta\right)\left(1-\cos \theta\right)^2}{2}\right]^{2/3} \right\} \neq 0 .
\end{eqnarray}
%Since there is little surface energy released, it becomes a puzzle for the occurrence of detachment. However, 
The actual numerical values of surface energy (area) change as a function of bubble radii for surface bubbles (for example, $\theta=150^{\circ}$, see the red dashed line) without using the limits are plotted in Fig.\,\ref{fig3}, again showing the decrease from approximately $20.0\%$ to nearly zero when reducing the bubble radius.

Obviously, the extremely small $E_s$ for 10 nm NBs cannot account for their jumping after merging. As shown in Fig.\,\ref{fig2}(f), $E_s^*$ reaches a small peak (of the value 0.05) at $t=0.1$ ns, corresponding to the flattening of the liquid bridge (State II) in Fig.\,\ref{fig2}(a). However, when the bubble velocity reaches its maximum at $t=0.3$~ns [Fig.\,\ref{fig2}(e)], $E_s^*$ has already returned to zero. This behavior contrasts sharply with droplet coalescence, where surface energy is released continuously (though with oscillations)\,\citep{liu2014numerical}, and suggests that another form of energy, previously unexplored, must play the dominant role.  

While the surface area change during bubble coalescence becomes negligible at the nanoscale, the bubble volume increases significantly. For free NBs, $V_2/(2V_1)=\sqrt{2}$, and for surface NBs, $V_2/(2V_1)=\sqrt{\left(2+\cos\theta\right)\left(1-\cos \theta\right)^2/2}$. These relations agree with the observed $\sim 40\%$ volume increase in Figs.\,\ref{fig2}(c) and 2(d). As the NB expands during coalescence, pressure energy (pressure-volume work) $E_p=\int_{V_0}^{V(t)} \left(P_g-P_{\infty}\right)\,dV$
is released\,\citep{oguz1993dynamics,brenner2002single}, where $P_g$ is the gas pressure and the initial volume $V_0=2V_1$. This pressure energy plays a central role in the Rayleigh–Plesset (RP) equation, which expresses the conservation of energy including surface energy, pressure energy, kinetic energy, and viscous dissipation (see SM-II). Similar to single-nanobubble dynamics described by the RP equation, the gas pressure inside the merged nanobubble is expected to remain uniform and follow some thermodynamic processes\,\citep{feng1997nonlinear,plesset1977bubble}. In our case, as the NB's volume oscillation is weak, the isothermal process is obeyed $P_gV_g=Nk_BT$\,\citep{prosperetti1988nonlinear}. Accordingly, the expression for pressure energy release during coalescence can be integrated as
\begin{equation}
   E_p=N_2k_BT\ln\frac{V_2}{2V_1}-P_{\infty}\left(V_2-2V_1\right). 
\end{equation}
In Fig.\,\ref{fig3}, the pressure energy (solid blue line for free NBs and dashed blue line for surface NBs with $\theta=150^{\circ}$) nondimensionalized by $8\pi \gamma R^2$ shows that it is negligible at the microscale. However, it becomes increasingly important when decreasing the bubble radius down to nanometers. Thus, there is an energy release transition from surface energy to pressure energy when reducing length scales. On the other hand, the total driving energy $E_{sp}$ 
\begin{equation}
    E_{sp}=E_s+E_p,
\end{equation}
shown by the black lines in Fig.\,\ref{fig3}, increases slightly when reducing length scales, providing sufficient driving energy for bubble detachment across all scales.
\begin{figure}[t]
\includegraphics [width=0.9\linewidth]{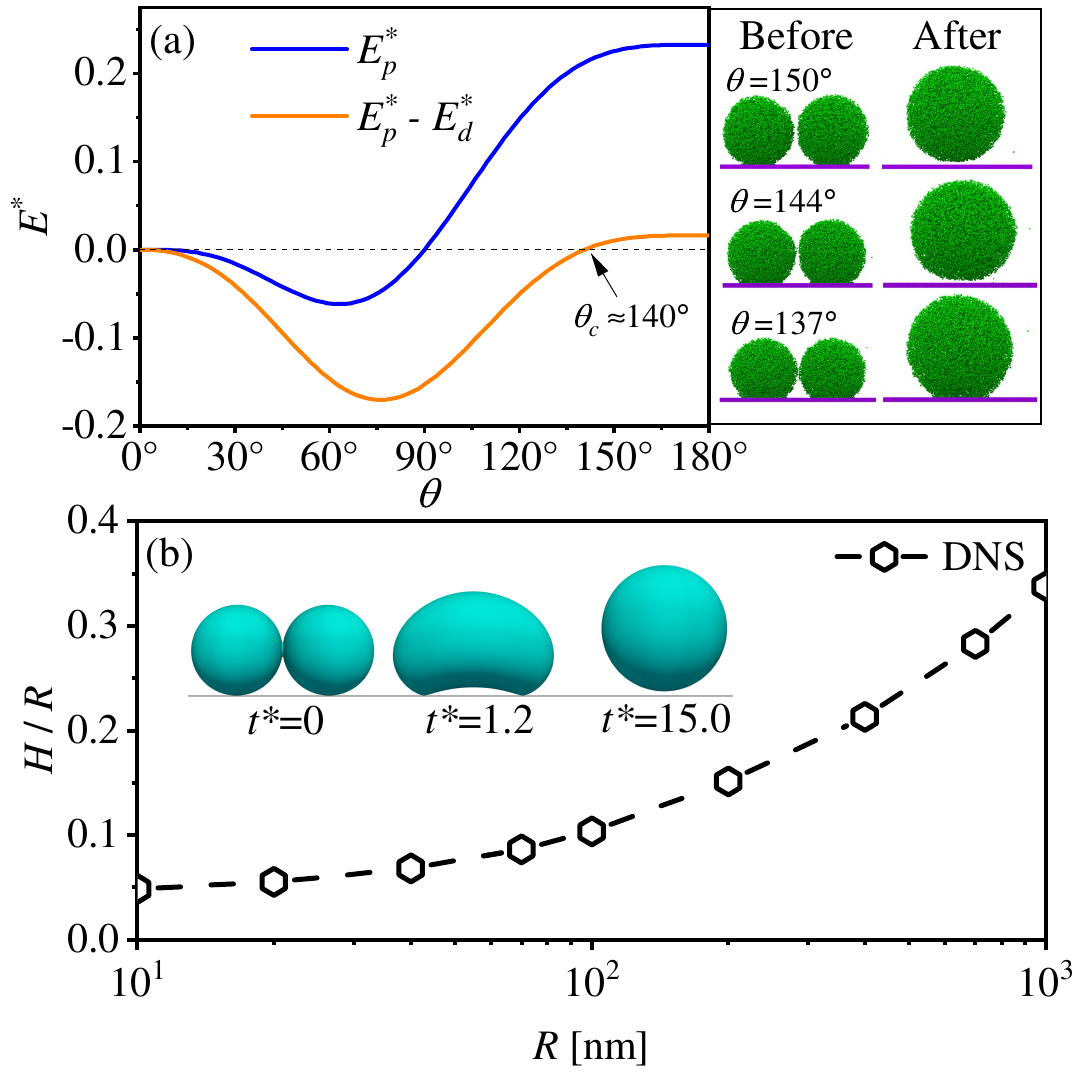}
\caption{\label{fig4} (a) Pressure energy $E_p$ as a function of contact angles with a fixed radius of curvature and the effects of viscous dissipation energy $E_{d}$. The MD snapshots given for three different initial contact angles $\theta$ show whether NBs can jump after coalescence. (b) Jumping distance $H$ (nondimensionalized by NB radii) as a function of initial NB radii, obtained by direct numerical simulations (DNS) of compressible NB coalescence. The inset pictures are simulation snapshots for 100 nm NBs with $\theta=180^{\circ}$, and $t^*=t/\sqrt{\rho_lR^3/\gamma}$.}
\end{figure}

Indeed, as shown by our MD results in Fig.\,\ref{fig2}(f), the pressure energy (blue line) dominates when the NB reaches its maximum velocity. After reaching the maximum, the NB velocity decreases and the NB starts to depart, and it completes departure at $t=0.78$ ns [State IV in Fig.\,\ref{fig2}(a)]. Similar to the droplet case, one can estimate the energy efficiency using the maximum jumping velocity\,\citep{liu2014numerical}. Theoretically, if all pressure energy turns into the bubble kinetic energy, $E_p=\left(\rho_l/2+\rho_g\right)Vu^2/2$, with ${\rho_lV}/{2}$ the added mass. This results in $u=99$ m/s. The maximum velocity from MD is, however, only about 8.7 m/s, which indicates an energy efficiency of only about $1\%$. Such a low efficiency is expected due to the high $Oh$ number, and it actually agrees with the coalescence of nanodroplets\,\citep{liu2014numerical,cheng2016numerical,perumanath2020molecular}, despite the driving energy being very different.

\emph{Critical contact angles for NB jumping and viscous dissipation}.\,---
As shown by the blue line in Fig.\,\ref{fig4}(a), $E_p$ for NBs decreases with decreasing contact angles (for $\theta>90^{\circ}$), making detachment increasingly difficult. This is similar to the relation between surface energy (area) change and contact angles for MBs' coalescence, i.e., Eq.\,\eqref{eq_dsmb}. Indeed, our simulations [see the snapshots in Fig.\,\ref{fig4}(a)] reveal that when the contact angle is reduced to about $137^{\circ}$, the merged nanobubble no longer detaches (also see Movie S2). This outcome is clearly also influenced by viscous dissipation. Strictly speaking, the exact value of viscous dissipation requires a detailed numerical computation\,\citep{cheng2016numerical,huang2019energy}. Here, we provide a simple estimate using the RP framework, since NBs' coalescence resembles the expansion of a compressed single NB. The viscous dissipation in the RP equation is given by (see SM-II)
\begin{equation}
    E_d=\int_{0}^{t} \int_{R}^{\infty} \Phi \, dV\,dt 
    = \int_{0}^{t} 16\pi \mu R \dot R^2 dt.
\end{equation}
Here $\Phi$ is the dissipation function. As oscillations are heavily damped by enhanced viscous effects at the nanoscale, the viscous damping timescale $\tau_{vis} = \rho_l R^2 / \mu$\,\citep{brennen2014cavitation} can be used to estimate $E_d \approx {16\pi \mu^2 R_2}/{\rho_l}$. Using a prefactor of $6.6$ instead of $16$ (see SM-II for explaining the deviation of the prefactor) leads to a critical angle of $140^{\circ}$ in reasonable agreement with MD results, as shown by the orange line in Fig.\,\ref{fig4}(a). The remaining pressure energy after subtracting viscous dissipation for $\theta=150^{\circ}$ is about $1\%$, which also again proves the low energy efficiency for jumping. In contrast, the classical viscous dissipation estimate $E_d\approx \mu \sqrt{\gamma/\rho_l }R_2^{3/2}$ for MBs based on capillary–inertial scaling\,\citep{lv2021self,demirkir2025jump,li2025size} severely underestimates dissipation (see SM-II).

\emph{Jumping distance of larger NBs and DNS simulations}.\,---
In our MD simulations, the jumping distance $H$ of a 10~nm NB after coalescence is about 0.7 nm, limited by the strong viscous drag. Although small, this displacement is sufficient to remove bubble blockage and may enhance the efficiency of many physicochemical processes, including water electrolysis. For larger NBs, one can expect greater jumping distances, as their deceleration after detachment decreases with increasing radii. Indeed, our MD simulations of 20 nm NBs show an increased jumping distance of 1.2 nm (see SM-I), requiring nearly 100 million atoms to eliminate the confinement of boundaries, at enormous computational cost. For even larger NBs, we employ computationally more efficient direct numerical simulations (DNS) to study \textit{compressible} bubble coalescence, since gas compressibility has been identified for the first time as the dominant effect governing nanobubble jumping. The DNS simulations are performed using the Basilisk code\,\citep{fuster2018all,popinet2018numerical} and the Basilisk results agree well with MD simulations (see SM-III) such as the minimal change of bubble area and the significant volume increase during coalescence, and confirm that the jumping distance increases with bubble radii, as shown in Fig.\,\ref{fig4}(b). Note that the jumping distance is nondimensionalized by the bubble radius. The current Basilisk simulation also only applies to bubbles with contact angles of $\theta=180^{\circ}$. 

\emph{Conclusions and Outlook}.\,--- Through molecular dynamics simulations, continuum numerical simulations, and theory, we reveal that the pressure energy release detaches the merged nanobubbles from substrates, compensating for the suppression of surface energy release at the nanoscale. By combining the pressure energy with the surface energy, our work thus unifies the driving mechanism for coalescence-induced gas bubble detachment across all length scales, which is crucial for enhancing the efficiency of many practical physicochemical processes involving gas evolution reactions. Previous experiments about nanobubble coalescence are limited for small contact angle nanobubbles\,\citep{nag2021dynamic,shin2015growth}. Our current work motivates new experiments about large contact angle nanobubble coalescence. Future work should also extend direct numerical simulations to include contact angle effects. Beyond detachment, gas compressibility may also influence other aspects of nanobubble coalescence, such as film drainage\,\citep{C0SM00812E,PhysRevLett.122.088002} and neck growth\,\citep{eggers2024coalescence}.
%For bubbles smaller than 10 nm, additional nanoscale effects, such as line tension, Tolman length, thermal fluctuations, and pressure-dependent interfacial energy, possibly play a role in their coalescence. 

\emph{Acknowledgment}.\,--- We are grateful for the discussions with J. Sprittles and P. Lv and acknowledge the financial support from the Advanced Research Center Chemical Building Blocks Consortium (ARC CBBC) under project (ARC CBBC 2021.038.UT.15). We are also grateful for the computational resources provided by the Dutch National supercomputer Snellius and EuroHPC supercomputer Discoverer.


\begin{thebibliography}{54}%
\makeatletter
\providecommand \@ifxundefined [1]{%
 \@ifx{#1\undefined}
}%
\providecommand \@ifnum [1]{%
 \ifnum #1\expandafter \@firstoftwo
 \else \expandafter \@secondoftwo
 \fi
}%
\providecommand \@ifx [1]{%
 \ifx #1\expandafter \@firstoftwo
 \else \expandafter \@secondoftwo
 \fi
}%
\providecommand \natexlab [1]{#1}%
\providecommand \enquote  [1]{``#1''}%
\providecommand \bibnamefont  [1]{#1}%
\providecommand \bibfnamefont [1]{#1}%
\providecommand \citenamefont [1]{#1}%
\providecommand \href@noop [0]{\@secondoftwo}%
\providecommand \href [0]{\begingroup \@sanitize@url \@href}%
\providecommand \@href[1]{\@@startlink{#1}\@@href}%
\providecommand \@@href[1]{\endgroup#1\@@endlink}%
\providecommand \@sanitize@url [0]{\catcode `\\12\catcode `\$12\catcode `\&12\catcode `\#12\catcode `\^12\catcode `\_12\catcode `\%12\relax}%
\providecommand \@@startlink[1]{}%
\providecommand \@@endlink[0]{}%
\providecommand \url  [0]{\begingroup\@sanitize@url \@url }%
\providecommand \@url [1]{\endgroup\@href {#1}{\urlprefix }}%
\providecommand \urlprefix  [0]{URL }%
\providecommand \Eprint [0]{\href }%
\providecommand \doibase [0]{https://doi.org/}%
\providecommand \selectlanguage [0]{\@gobble}%
\providecommand \bibinfo  [0]{\@secondoftwo}%
\providecommand \bibfield  [0]{\@secondoftwo}%
\providecommand \translation [1]{[#1]}%
\providecommand \BibitemOpen [0]{}%
\providecommand \bibitemStop [0]{}%
\providecommand \bibitemNoStop [0]{.\EOS\space}%
\providecommand \EOS [0]{\spacefactor3000\relax}%
\providecommand \BibitemShut  [1]{\csname bibitem#1\endcsname}%
\let\auto@bib@innerbib\@empty
%</preamble>
\bibitem [{\citenamefont {Lohse}(2018)}]{lohse2018bubble}%
  \BibitemOpen
  \bibfield  {author} {\bibinfo {author} {\bibfnamefont {D.}~\bibnamefont {Lohse}},\ }\bibfield  {title} {\bibinfo {title} {Bubble puzzles: From fundamentals to applications},\ }\href@noop {} {\bibfield  {journal} {\bibinfo  {journal} {Phys. Rev. Fluids}\ }\textbf {\bibinfo {volume} {3}},\ \bibinfo {pages} {110504} (\bibinfo {year} {2018})}\BibitemShut {NoStop}%
\bibitem [{\citenamefont {Bashkatov}\ \emph {et~al.}(2019)\citenamefont {Bashkatov}, \citenamefont {Hossain}, \citenamefont {Yang}, \citenamefont {Mutschke},\ and\ \citenamefont {Eckert}}]{PhysRevLett.123.214503}%
  \BibitemOpen
  \bibfield  {author} {\bibinfo {author} {\bibfnamefont {A.}~\bibnamefont {Bashkatov}}, \bibinfo {author} {\bibfnamefont {S.~S.}\ \bibnamefont {Hossain}}, \bibinfo {author} {\bibfnamefont {X.}~\bibnamefont {Yang}}, \bibinfo {author} {\bibfnamefont {G.}~\bibnamefont {Mutschke}},\ and\ \bibinfo {author} {\bibfnamefont {K.}~\bibnamefont {Eckert}},\ }\bibfield  {title} {\bibinfo {title} {Oscillating hydrogen bubbles at {Pt} microelectrodes},\ }\href@noop {} {\bibfield  {journal} {\bibinfo  {journal} {Phys. Rev. Lett.}\ }\textbf {\bibinfo {volume} {123}},\ \bibinfo {pages} {214503} (\bibinfo {year} {2019})}\BibitemShut {NoStop}%
\bibitem [{\citenamefont {Zhang}\ \emph {et~al.}(2024)\citenamefont {Zhang}, \citenamefont {Zhu}, \citenamefont {Wood},\ and\ \citenamefont {Lohse}}]{zhang2024threshold}%
  \BibitemOpen
  \bibfield  {author} {\bibinfo {author} {\bibfnamefont {Y.}~\bibnamefont {Zhang}}, \bibinfo {author} {\bibfnamefont {X.}~\bibnamefont {Zhu}}, \bibinfo {author} {\bibfnamefont {J.~A.}\ \bibnamefont {Wood}},\ and\ \bibinfo {author} {\bibfnamefont {D.}~\bibnamefont {Lohse}},\ }\bibfield  {title} {\bibinfo {title} {Threshold current density for diffusion-controlled stability of electrolytic surface nanobubbles},\ }\href@noop {} {\bibfield  {journal} {\bibinfo  {journal} {Proc. Natl. Acad. Sci. U.S.A.}\ }\textbf {\bibinfo {volume} {121}},\ \bibinfo {pages} {e2321958121} (\bibinfo {year} {2024})}\BibitemShut {NoStop}%
\bibitem [{\citenamefont {Angulo}\ \emph {et~al.}(2020)\citenamefont {Angulo}, \citenamefont {van~der Linde}, \citenamefont {Gardeniers}, \citenamefont {Modestino},\ and\ \citenamefont {Rivas}}]{angulo2020influence}%
  \BibitemOpen
  \bibfield  {author} {\bibinfo {author} {\bibfnamefont {A.}~\bibnamefont {Angulo}}, \bibinfo {author} {\bibfnamefont {P.}~\bibnamefont {van~der Linde}}, \bibinfo {author} {\bibfnamefont {H.}~\bibnamefont {Gardeniers}}, \bibinfo {author} {\bibfnamefont {M.}~\bibnamefont {Modestino}},\ and\ \bibinfo {author} {\bibfnamefont {D.~F.}\ \bibnamefont {Rivas}},\ }\bibfield  {title} {\bibinfo {title} {Influence of bubbles on the energy conversion efficiency of electrochemical reactors},\ }\href@noop {} {\bibfield  {journal} {\bibinfo  {journal} {Joule}\ }\textbf {\bibinfo {volume} {4}},\ \bibinfo {pages} {555} (\bibinfo {year} {2020})}\BibitemShut {NoStop}%
\bibitem [{\citenamefont {Bashkatov}\ \emph {et~al.}(2024)\citenamefont {Bashkatov}, \citenamefont {Park}, \citenamefont {Demirk{\i}r}, \citenamefont {Wood}, \citenamefont {Koper}, \citenamefont {Lohse},\ and\ \citenamefont {Krug}}]{bashkatov2024performance}%
  \BibitemOpen
  \bibfield  {author} {\bibinfo {author} {\bibfnamefont {A.}~\bibnamefont {Bashkatov}}, \bibinfo {author} {\bibfnamefont {S.}~\bibnamefont {Park}}, \bibinfo {author} {\bibfnamefont {{\c{C}}.}~\bibnamefont {Demirk{\i}r}}, \bibinfo {author} {\bibfnamefont {J.~A.}\ \bibnamefont {Wood}}, \bibinfo {author} {\bibfnamefont {M.~T.}\ \bibnamefont {Koper}}, \bibinfo {author} {\bibfnamefont {D.}~\bibnamefont {Lohse}},\ and\ \bibinfo {author} {\bibfnamefont {D.}~\bibnamefont {Krug}},\ }\bibfield  {title} {\bibinfo {title} {Performance enhancement of electrocatalytic hydrogen evolution through coalescence-induced bubble dynamics},\ }\href@noop {} {\bibfield  {journal} {\bibinfo  {journal} {J. Am. Chem. Soc.}\ }\textbf {\bibinfo {volume} {146}},\ \bibinfo {pages} {10177} (\bibinfo {year} {2024})}\BibitemShut {NoStop}%
\bibitem [{\citenamefont {Zhao}\ \emph {et~al.}(2019)\citenamefont {Zhao}, \citenamefont {Ren},\ and\ \citenamefont {Luo}}]{zhao2019gas}%
  \BibitemOpen
  \bibfield  {author} {\bibinfo {author} {\bibfnamefont {X.}~\bibnamefont {Zhao}}, \bibinfo {author} {\bibfnamefont {H.}~\bibnamefont {Ren}},\ and\ \bibinfo {author} {\bibfnamefont {L.}~\bibnamefont {Luo}},\ }\bibfield  {title} {\bibinfo {title} {Gas bubbles in electrochemical gas evolution reactions},\ }\href@noop {} {\bibfield  {journal} {\bibinfo  {journal} {Langmuir}\ }\textbf {\bibinfo {volume} {35}},\ \bibinfo {pages} {5392} (\bibinfo {year} {2019})}\BibitemShut {NoStop}%
\bibitem [{\citenamefont {Shih}\ \emph {et~al.}(2022)\citenamefont {Shih}, \citenamefont {Monteiro}, \citenamefont {Dattila}, \citenamefont {Pavesi}, \citenamefont {Philips}, \citenamefont {da~Silva}, \citenamefont {Vos}, \citenamefont {Ojha}, \citenamefont {Park}, \citenamefont {van~der Heijden} \emph {et~al.}}]{shih2022water}%
  \BibitemOpen
  \bibfield  {author} {\bibinfo {author} {\bibfnamefont {A.~J.}\ \bibnamefont {Shih}}, \bibinfo {author} {\bibfnamefont {M.~C.}\ \bibnamefont {Monteiro}}, \bibinfo {author} {\bibfnamefont {F.}~\bibnamefont {Dattila}}, \bibinfo {author} {\bibfnamefont {D.}~\bibnamefont {Pavesi}}, \bibinfo {author} {\bibfnamefont {M.}~\bibnamefont {Philips}}, \bibinfo {author} {\bibfnamefont {A.~H.}\ \bibnamefont {da~Silva}}, \bibinfo {author} {\bibfnamefont {R.~E.}\ \bibnamefont {Vos}}, \bibinfo {author} {\bibfnamefont {K.}~\bibnamefont {Ojha}}, \bibinfo {author} {\bibfnamefont {S.}~\bibnamefont {Park}}, \bibinfo {author} {\bibfnamefont {O.}~\bibnamefont {van~der Heijden}}, \emph {et~al.},\ }\bibfield  {title} {\bibinfo {title} {Water electrolysis},\ }\href@noop {} {\bibfield  {journal} {\bibinfo  {journal} {Nat. Rev. Methods Primers.}\ }\textbf {\bibinfo {volume} {2}},\ \bibinfo {pages} {84} (\bibinfo {year} {2022})}\BibitemShut {NoStop}%
\bibitem [{\citenamefont {Li}\ \emph {et~al.}(2021)\citenamefont {Li}, \citenamefont {Kang}, \citenamefont {Fazle~Rabbi}, \citenamefont {Fu}, \citenamefont {Yan}, \citenamefont {Fang}, \citenamefont {Fan},\ and\ \citenamefont {Miljkovic}}]{li2021liquid}%
  \BibitemOpen
  \bibfield  {author} {\bibinfo {author} {\bibfnamefont {J.}~\bibnamefont {Li}}, \bibinfo {author} {\bibfnamefont {D.}~\bibnamefont {Kang}}, \bibinfo {author} {\bibfnamefont {K.}~\bibnamefont {Fazle~Rabbi}}, \bibinfo {author} {\bibfnamefont {W.}~\bibnamefont {Fu}}, \bibinfo {author} {\bibfnamefont {X.}~\bibnamefont {Yan}}, \bibinfo {author} {\bibfnamefont {X.}~\bibnamefont {Fang}}, \bibinfo {author} {\bibfnamefont {L.}~\bibnamefont {Fan}},\ and\ \bibinfo {author} {\bibfnamefont {N.}~\bibnamefont {Miljkovic}},\ }\bibfield  {title} {\bibinfo {title} {Liquid film--induced critical heat flux enhancement on structured surfaces},\ }\href@noop {} {\bibfield  {journal} {\bibinfo  {journal} {Sci. Adv.}\ }\textbf {\bibinfo {volume} {7}},\ \bibinfo {pages} {eabg4537} (\bibinfo {year} {2021})}\BibitemShut {NoStop}%
\bibitem [{\citenamefont {Gallo}\ \emph {et~al.}(2023)\citenamefont {Gallo}, \citenamefont {Magaletti}, \citenamefont {Georgoulas}, \citenamefont {Marengo}, \citenamefont {De~Coninck},\ and\ \citenamefont {Casciola}}]{gallo2023nanoscale}%
  \BibitemOpen
  \bibfield  {author} {\bibinfo {author} {\bibfnamefont {M.}~\bibnamefont {Gallo}}, \bibinfo {author} {\bibfnamefont {F.}~\bibnamefont {Magaletti}}, \bibinfo {author} {\bibfnamefont {A.}~\bibnamefont {Georgoulas}}, \bibinfo {author} {\bibfnamefont {M.}~\bibnamefont {Marengo}}, \bibinfo {author} {\bibfnamefont {J.}~\bibnamefont {De~Coninck}},\ and\ \bibinfo {author} {\bibfnamefont {C.~M.}\ \bibnamefont {Casciola}},\ }\bibfield  {title} {\bibinfo {title} {A nanoscale view of the origin of boiling and its dynamics},\ }\href@noop {} {\bibfield  {journal} {\bibinfo  {journal} {Nat. Commun.}\ }\textbf {\bibinfo {volume} {14}},\ \bibinfo {pages} {6428} (\bibinfo {year} {2023})}\BibitemShut {NoStop}%
\bibitem [{\citenamefont {Park}\ \emph {et~al.}(2024)\citenamefont {Park}, \citenamefont {Ahmadi}, \citenamefont {Foulkes},\ and\ \citenamefont {Boreyko}}]{park2024coalescence}%
  \BibitemOpen
  \bibfield  {author} {\bibinfo {author} {\bibfnamefont {H.}~\bibnamefont {Park}}, \bibinfo {author} {\bibfnamefont {S.~F.}\ \bibnamefont {Ahmadi}}, \bibinfo {author} {\bibfnamefont {T.~P.}\ \bibnamefont {Foulkes}},\ and\ \bibinfo {author} {\bibfnamefont {J.~B.}\ \bibnamefont {Boreyko}},\ }\bibfield  {title} {\bibinfo {title} {Coalescence-induced jumping bubbles during pool boiling},\ }\href@noop {} {\bibfield  {journal} {\bibinfo  {journal} {Adv. Funct. Mater.}\ }\textbf {\bibinfo {volume} {34}},\ \bibinfo {pages} {2312088} (\bibinfo {year} {2024})}\BibitemShut {NoStop}%
\bibitem [{\citenamefont {Prosperetti}(2017)}]{prosperetti2017vapor}%
  \BibitemOpen
  \bibfield  {author} {\bibinfo {author} {\bibfnamefont {A.}~\bibnamefont {Prosperetti}},\ }\bibfield  {title} {\bibinfo {title} {Vapor bubbles},\ }\href@noop {} {\bibfield  {journal} {\bibinfo  {journal} {Annu. Rev. Fluid Mech.}\ }\textbf {\bibinfo {volume} {49}},\ \bibinfo {pages} {221} (\bibinfo {year} {2017})}\BibitemShut {NoStop}%
\bibitem [{\citenamefont {Kavokine}\ \emph {et~al.}(2021)\citenamefont {Kavokine}, \citenamefont {Netz},\ and\ \citenamefont {Bocquet}}]{kavokine2021fluids}%
  \BibitemOpen
  \bibfield  {author} {\bibinfo {author} {\bibfnamefont {N.}~\bibnamefont {Kavokine}}, \bibinfo {author} {\bibfnamefont {R.~R.}\ \bibnamefont {Netz}},\ and\ \bibinfo {author} {\bibfnamefont {L.}~\bibnamefont {Bocquet}},\ }\bibfield  {title} {\bibinfo {title} {Fluids at the nanoscale: From continuum to subcontinuum transport},\ }\href@noop {} {\bibfield  {journal} {\bibinfo  {journal} {Annu. Rev. Fluid Mech.}\ }\textbf {\bibinfo {volume} {53}},\ \bibinfo {pages} {377} (\bibinfo {year} {2021})}\BibitemShut {NoStop}%
\bibitem [{\citenamefont {Singh}\ \emph {et~al.}(2015)\citenamefont {Singh}, \citenamefont {Clark},\ and\ \citenamefont {Bell}}]{singh2015effects}%
  \BibitemOpen
  \bibfield  {author} {\bibinfo {author} {\bibfnamefont {M.~R.}\ \bibnamefont {Singh}}, \bibinfo {author} {\bibfnamefont {E.~L.}\ \bibnamefont {Clark}},\ and\ \bibinfo {author} {\bibfnamefont {A.~T.}\ \bibnamefont {Bell}},\ }\bibfield  {title} {\bibinfo {title} {Effects of electrolyte, catalyst, and membrane composition and operating conditions on the performance of solar-driven electrochemical reduction of carbon dioxide},\ }\href@noop {} {\bibfield  {journal} {\bibinfo  {journal} {Phys. Chem. Chem. Phys.}\ }\textbf {\bibinfo {volume} {17}},\ \bibinfo {pages} {18924} (\bibinfo {year} {2015})}\BibitemShut {NoStop}%
\bibitem [{\citenamefont {He}\ \emph {et~al.}(2023)\citenamefont {He}, \citenamefont {Cui}, \citenamefont {Zhao}, \citenamefont {Chen}, \citenamefont {Shang},\ and\ \citenamefont {Tan}}]{he2023strategies}%
  \BibitemOpen
  \bibfield  {author} {\bibinfo {author} {\bibfnamefont {Y.}~\bibnamefont {He}}, \bibinfo {author} {\bibfnamefont {Y.}~\bibnamefont {Cui}}, \bibinfo {author} {\bibfnamefont {Z.}~\bibnamefont {Zhao}}, \bibinfo {author} {\bibfnamefont {Y.}~\bibnamefont {Chen}}, \bibinfo {author} {\bibfnamefont {W.}~\bibnamefont {Shang}},\ and\ \bibinfo {author} {\bibfnamefont {P.}~\bibnamefont {Tan}},\ }\bibfield  {title} {\bibinfo {title} {Strategies for bubble removal in electrochemical systems},\ }\href@noop {} {\bibfield  {journal} {\bibinfo  {journal} {Energy Rev.}\ }\textbf {\bibinfo {volume} {2}},\ \bibinfo {pages} {100015} (\bibinfo {year} {2023})}\BibitemShut {NoStop}%
\bibitem [{\citenamefont {Soto}\ \emph {et~al.}(2018)\citenamefont {Soto}, \citenamefont {Maddalena}, \citenamefont {Fraters}, \citenamefont {Van Der~Meer},\ and\ \citenamefont {Lohse}}]{soto2018coalescence}%
  \BibitemOpen
  \bibfield  {author} {\bibinfo {author} {\bibfnamefont {{\'A}.~M.}\ \bibnamefont {Soto}}, \bibinfo {author} {\bibfnamefont {T.}~\bibnamefont {Maddalena}}, \bibinfo {author} {\bibfnamefont {A.}~\bibnamefont {Fraters}}, \bibinfo {author} {\bibfnamefont {D.}~\bibnamefont {Van Der~Meer}},\ and\ \bibinfo {author} {\bibfnamefont {D.}~\bibnamefont {Lohse}},\ }\bibfield  {title} {\bibinfo {title} {Coalescence of diffusively growing gas bubbles},\ }\href@noop {} {\bibfield  {journal} {\bibinfo  {journal} {J. Fluid Mech.}\ }\textbf {\bibinfo {volume} {846}},\ \bibinfo {pages} {143} (\bibinfo {year} {2018})}\BibitemShut {NoStop}%
\bibitem [{\citenamefont {Lv}\ \emph {et~al.}(2021)\citenamefont {Lv}, \citenamefont {Pe{\~n}as}, \citenamefont {Le~The}, \citenamefont {Eijkel}, \citenamefont {van~den Berg}, \citenamefont {Zhang},\ and\ \citenamefont {Lohse}}]{lv2021self}%
  \BibitemOpen
  \bibfield  {author} {\bibinfo {author} {\bibfnamefont {P.}~\bibnamefont {Lv}}, \bibinfo {author} {\bibfnamefont {P.}~\bibnamefont {Pe{\~n}as}}, \bibinfo {author} {\bibfnamefont {H.}~\bibnamefont {Le~The}}, \bibinfo {author} {\bibfnamefont {J.}~\bibnamefont {Eijkel}}, \bibinfo {author} {\bibfnamefont {A.}~\bibnamefont {van~den Berg}}, \bibinfo {author} {\bibfnamefont {X.}~\bibnamefont {Zhang}},\ and\ \bibinfo {author} {\bibfnamefont {D.}~\bibnamefont {Lohse}},\ }\bibfield  {title} {\bibinfo {title} {Self-propelled detachment upon coalescence of surface bubbles},\ }\href@noop {} {\bibfield  {journal} {\bibinfo  {journal} {Phys. Rev. Lett.}\ }\textbf {\bibinfo {volume} {127}},\ \bibinfo {pages} {235501} (\bibinfo {year} {2021})}\BibitemShut {NoStop}%
\bibitem [{\citenamefont {Li}\ \emph {et~al.}(2025)\citenamefont {Li}, \citenamefont {Xu}, \citenamefont {Luo}, \citenamefont {Nie}, \citenamefont {Wang}, \citenamefont {She},\ and\ \citenamefont {Guo}}]{li2025size}%
  \BibitemOpen
  \bibfield  {author} {\bibinfo {author} {\bibfnamefont {J.}~\bibnamefont {Li}}, \bibinfo {author} {\bibfnamefont {Q.}~\bibnamefont {Xu}}, \bibinfo {author} {\bibfnamefont {X.}~\bibnamefont {Luo}}, \bibinfo {author} {\bibfnamefont {T.}~\bibnamefont {Nie}}, \bibinfo {author} {\bibfnamefont {M.}~\bibnamefont {Wang}}, \bibinfo {author} {\bibfnamefont {Y.}~\bibnamefont {She}},\ and\ \bibinfo {author} {\bibfnamefont {L.}~\bibnamefont {Guo}},\ }\bibfield  {title} {\bibinfo {title} {Size effects on bubble dynamics during photoelectrochemical water splitting},\ }\href@noop {} {\bibfield  {journal} {\bibinfo  {journal} {ACS Nano}\ }\textbf {\bibinfo {volume} {19}},\ \bibinfo {pages} {8200} (\bibinfo {year} {2025})}\BibitemShut {NoStop}%
\bibitem [{\citenamefont {Çayan Demirkır}\ \emph {et~al.}(2025)\citenamefont {Çayan Demirkır}, \citenamefont {Yang}, \citenamefont {Bashkatov}, \citenamefont {Sanjay}, \citenamefont {Lohse},\ and\ \citenamefont {Krug}}]{demirkir2025jump}%
  \BibitemOpen
  \bibfield  {author} {\bibinfo {author} {\bibnamefont {Çayan Demirkır}}, \bibinfo {author} {\bibfnamefont {R.}~\bibnamefont {Yang}}, \bibinfo {author} {\bibfnamefont {A.}~\bibnamefont {Bashkatov}}, \bibinfo {author} {\bibfnamefont {V.}~\bibnamefont {Sanjay}}, \bibinfo {author} {\bibfnamefont {D.}~\bibnamefont {Lohse}},\ and\ \bibinfo {author} {\bibfnamefont {D.}~\bibnamefont {Krug}},\ }\href {https://arxiv.org/abs/2501.05532} {\bibinfo {title} {To jump or not to jump: Adhesion and viscous dissipation dictate the detachment of coalescing wall-attached bubbles}} (\bibinfo {year} {2025}),\ \Eprint {https://arxiv.org/abs/2501.05532} {arXiv:2501.05532 [physics.flu-dyn]} \BibitemShut {NoStop}%
\bibitem [{\citenamefont {Lu}\ \emph {et~al.}(2025)\citenamefont {Lu}, \citenamefont {Yadav}, \citenamefont {Zhou}, \citenamefont {Zhou}, \citenamefont {Liu}, \citenamefont {Li}, \citenamefont {Ma},\ and\ \citenamefont {Jing}}]{lu2025combined}%
  \BibitemOpen
  \bibfield  {author} {\bibinfo {author} {\bibfnamefont {X.}~\bibnamefont {Lu}}, \bibinfo {author} {\bibfnamefont {D.}~\bibnamefont {Yadav}}, \bibinfo {author} {\bibfnamefont {L.}~\bibnamefont {Zhou}}, \bibinfo {author} {\bibfnamefont {Y.}~\bibnamefont {Zhou}}, \bibinfo {author} {\bibfnamefont {Q.}~\bibnamefont {Liu}}, \bibinfo {author} {\bibfnamefont {X.}~\bibnamefont {Li}}, \bibinfo {author} {\bibfnamefont {L.}~\bibnamefont {Ma}},\ and\ \bibinfo {author} {\bibfnamefont {D.}~\bibnamefont {Jing}},\ }\bibfield  {title} {\bibinfo {title} {Combined effects of bubble size ratio on coalescence-driven dynamics and electrode dimensions on electrolytic performance},\ }\href@noop {} {\bibfield  {journal} {\bibinfo  {journal} {Langmuir}\ }\textbf {\bibinfo {volume} {41}},\ \bibinfo {pages} {20917} (\bibinfo {year} {2025})}\BibitemShut {NoStop}%
\bibitem [{\citenamefont {Zhao}\ \emph {et~al.}(2022)\citenamefont {Zhao}, \citenamefont {Hu}, \citenamefont {Cheng}, \citenamefont {Huang},\ and\ \citenamefont {Gong}}]{zhao2022coalescence}%
  \BibitemOpen
  \bibfield  {author} {\bibinfo {author} {\bibfnamefont {P.}~\bibnamefont {Zhao}}, \bibinfo {author} {\bibfnamefont {Z.}~\bibnamefont {Hu}}, \bibinfo {author} {\bibfnamefont {P.}~\bibnamefont {Cheng}}, \bibinfo {author} {\bibfnamefont {R.}~\bibnamefont {Huang}},\ and\ \bibinfo {author} {\bibfnamefont {S.}~\bibnamefont {Gong}},\ }\bibfield  {title} {\bibinfo {title} {Coalescence-induced bubble departure: {Effects} of dynamic contact angles},\ }\href@noop {} {\bibfield  {journal} {\bibinfo  {journal} {Langmuir}\ }\textbf {\bibinfo {volume} {38}},\ \bibinfo {pages} {10558} (\bibinfo {year} {2022})}\BibitemShut {NoStop}%
\bibitem [{\citenamefont {Iwata}\ \emph {et~al.}(2022)\citenamefont {Iwata}, \citenamefont {Zhang}, \citenamefont {Lu}, \citenamefont {Gong}, \citenamefont {Du},\ and\ \citenamefont {Wang}}]{iwata2022coalescing}%
  \BibitemOpen
  \bibfield  {author} {\bibinfo {author} {\bibfnamefont {R.}~\bibnamefont {Iwata}}, \bibinfo {author} {\bibfnamefont {L.}~\bibnamefont {Zhang}}, \bibinfo {author} {\bibfnamefont {Z.}~\bibnamefont {Lu}}, \bibinfo {author} {\bibfnamefont {S.}~\bibnamefont {Gong}}, \bibinfo {author} {\bibfnamefont {J.}~\bibnamefont {Du}},\ and\ \bibinfo {author} {\bibfnamefont {E.~N.}\ \bibnamefont {Wang}},\ }\bibfield  {title} {\bibinfo {title} {How coalescing bubbles depart from a wall},\ }\href@noop {} {\bibfield  {journal} {\bibinfo  {journal} {Langmuir}\ }\textbf {\bibinfo {volume} {38}},\ \bibinfo {pages} {4371} (\bibinfo {year} {2022})}\BibitemShut {NoStop}%
\bibitem [{\citenamefont {Boreyko}\ and\ \citenamefont {Chen}(2009)}]{boreyko2009self}%
  \BibitemOpen
  \bibfield  {author} {\bibinfo {author} {\bibfnamefont {J.~B.}\ \bibnamefont {Boreyko}}\ and\ \bibinfo {author} {\bibfnamefont {C.-H.}\ \bibnamefont {Chen}},\ }\bibfield  {title} {\bibinfo {title} {Self-propelled dropwise condensate on superhydrophobic surfaces},\ }\href@noop {} {\bibfield  {journal} {\bibinfo  {journal} {Phys. Rev. Lett.}\ }\textbf {\bibinfo {volume} {103}},\ \bibinfo {pages} {184501} (\bibinfo {year} {2009})}\BibitemShut {NoStop}%
\bibitem [{\citenamefont {Boreyko}(2024)}]{boreyko2024jumping}%
  \BibitemOpen
  \bibfield  {author} {\bibinfo {author} {\bibfnamefont {J.~B.}\ \bibnamefont {Boreyko}},\ }\bibfield  {title} {\bibinfo {title} {Jumping droplets},\ }\href@noop {} {\bibfield  {journal} {\bibinfo  {journal} {Droplet}\ }\textbf {\bibinfo {volume} {3}},\ \bibinfo {pages} {e105} (\bibinfo {year} {2024})}\BibitemShut {NoStop}%
\bibitem [{\citenamefont {Aarts}\ \emph {et~al.}(2005)\citenamefont {Aarts}, \citenamefont {Lekkerkerker}, \citenamefont {Guo}, \citenamefont {Wegdam},\ and\ \citenamefont {Bonn}}]{PhysRevLett.95.164503}%
  \BibitemOpen
  \bibfield  {author} {\bibinfo {author} {\bibfnamefont {D.~G. A.~L.}\ \bibnamefont {Aarts}}, \bibinfo {author} {\bibfnamefont {H.~N.~W.}\ \bibnamefont {Lekkerkerker}}, \bibinfo {author} {\bibfnamefont {H.}~\bibnamefont {Guo}}, \bibinfo {author} {\bibfnamefont {G.~H.}\ \bibnamefont {Wegdam}},\ and\ \bibinfo {author} {\bibfnamefont {D.}~\bibnamefont {Bonn}},\ }\bibfield  {title} {\bibinfo {title} {Hydrodynamics of droplet coalescence},\ }\href@noop {} {\bibfield  {journal} {\bibinfo  {journal} {Phys. Rev. Lett.}\ }\textbf {\bibinfo {volume} {95}},\ \bibinfo {pages} {164503} (\bibinfo {year} {2005})}\BibitemShut {NoStop}%
\bibitem [{\citenamefont {Lv}\ \emph {et~al.}(2013)\citenamefont {Lv}, \citenamefont {Hao}, \citenamefont {Yao}, \citenamefont {Song}, \citenamefont {Zhang},\ and\ \citenamefont {He}}]{lv2013condensation}%
  \BibitemOpen
  \bibfield  {author} {\bibinfo {author} {\bibfnamefont {C.}~\bibnamefont {Lv}}, \bibinfo {author} {\bibfnamefont {P.}~\bibnamefont {Hao}}, \bibinfo {author} {\bibfnamefont {Z.}~\bibnamefont {Yao}}, \bibinfo {author} {\bibfnamefont {Y.}~\bibnamefont {Song}}, \bibinfo {author} {\bibfnamefont {X.}~\bibnamefont {Zhang}},\ and\ \bibinfo {author} {\bibfnamefont {F.}~\bibnamefont {He}},\ }\bibfield  {title} {\bibinfo {title} {Condensation and jumping relay of droplets on lotus leaf},\ }\href@noop {} {\bibfield  {journal} {\bibinfo  {journal} {Appl. Phys. Lett.}\ }\textbf {\bibinfo {volume} {103}} (\bibinfo {year} {2013})}\BibitemShut {NoStop}%
\bibitem [{\citenamefont {Enright}\ \emph {et~al.}(2014)\citenamefont {Enright}, \citenamefont {Miljkovic}, \citenamefont {Sprittles}, \citenamefont {Nolan}, \citenamefont {Mitchell},\ and\ \citenamefont {Wang}}]{enright2014coalescing}%
  \BibitemOpen
  \bibfield  {author} {\bibinfo {author} {\bibfnamefont {R.}~\bibnamefont {Enright}}, \bibinfo {author} {\bibfnamefont {N.}~\bibnamefont {Miljkovic}}, \bibinfo {author} {\bibfnamefont {J.}~\bibnamefont {Sprittles}}, \bibinfo {author} {\bibfnamefont {K.}~\bibnamefont {Nolan}}, \bibinfo {author} {\bibfnamefont {R.}~\bibnamefont {Mitchell}},\ and\ \bibinfo {author} {\bibfnamefont {E.~N.}\ \bibnamefont {Wang}},\ }\bibfield  {title} {\bibinfo {title} {How coalescing droplets jump},\ }\href@noop {} {\bibfield  {journal} {\bibinfo  {journal} {ACS Nano}\ }\textbf {\bibinfo {volume} {8}},\ \bibinfo {pages} {10352} (\bibinfo {year} {2014})}\BibitemShut {NoStop}%
\bibitem [{\citenamefont {Liu}\ \emph {et~al.}(2014)\citenamefont {Liu}, \citenamefont {Ghigliotti}, \citenamefont {Feng},\ and\ \citenamefont {Chen}}]{liu2014numerical}%
  \BibitemOpen
  \bibfield  {author} {\bibinfo {author} {\bibfnamefont {F.}~\bibnamefont {Liu}}, \bibinfo {author} {\bibfnamefont {G.}~\bibnamefont {Ghigliotti}}, \bibinfo {author} {\bibfnamefont {J.~J.}\ \bibnamefont {Feng}},\ and\ \bibinfo {author} {\bibfnamefont {C.-H.}\ \bibnamefont {Chen}},\ }\bibfield  {title} {\bibinfo {title} {Numerical simulations of self-propelled jumping upon drop coalescence on non-wetting surfaces},\ }\href@noop {} {\bibfield  {journal} {\bibinfo  {journal} {J. Fluid Mech.}\ }\textbf {\bibinfo {volume} {752}},\ \bibinfo {pages} {39} (\bibinfo {year} {2014})}\BibitemShut {NoStop}%
\bibitem [{\citenamefont {Mouterde}\ \emph {et~al.}(2017)\citenamefont {Mouterde}, \citenamefont {Nguyen}, \citenamefont {Takahashi}, \citenamefont {Clanet}, \citenamefont {Shimoyama},\ and\ \citenamefont {Qu{\'e}r{\'e}}}]{mouterde2017merging}%
  \BibitemOpen
  \bibfield  {author} {\bibinfo {author} {\bibfnamefont {T.}~\bibnamefont {Mouterde}}, \bibinfo {author} {\bibfnamefont {T.-V.}\ \bibnamefont {Nguyen}}, \bibinfo {author} {\bibfnamefont {H.}~\bibnamefont {Takahashi}}, \bibinfo {author} {\bibfnamefont {C.}~\bibnamefont {Clanet}}, \bibinfo {author} {\bibfnamefont {I.}~\bibnamefont {Shimoyama}},\ and\ \bibinfo {author} {\bibfnamefont {D.}~\bibnamefont {Qu{\'e}r{\'e}}},\ }\bibfield  {title} {\bibinfo {title} {How merging droplets jump off a superhydrophobic surface: Measurements and model},\ }\href@noop {} {\bibfield  {journal} {\bibinfo  {journal} {Phys. Rev. Fluids}\ }\textbf {\bibinfo {volume} {2}},\ \bibinfo {pages} {112001} (\bibinfo {year} {2017})}\BibitemShut {NoStop}%
\bibitem [{\citenamefont {Vo}\ and\ \citenamefont {Tran}(2019)}]{vo2019critical}%
  \BibitemOpen
  \bibfield  {author} {\bibinfo {author} {\bibfnamefont {Q.}~\bibnamefont {Vo}}\ and\ \bibinfo {author} {\bibfnamefont {T.}~\bibnamefont {Tran}},\ }\bibfield  {title} {\bibinfo {title} {Critical conditions for jumping droplets},\ }\href@noop {} {\bibfield  {journal} {\bibinfo  {journal} {Phys. Rev. Lett.}\ }\textbf {\bibinfo {volume} {123}},\ \bibinfo {pages} {024502} (\bibinfo {year} {2019})}\BibitemShut {NoStop}%
\bibitem [{\citenamefont {Bird}\ \emph {et~al.}(2021)\citenamefont {Bird}, \citenamefont {Smith},\ and\ \citenamefont {Liang}}]{bird2021coalescence}%
  \BibitemOpen
  \bibfield  {author} {\bibinfo {author} {\bibfnamefont {E.}~\bibnamefont {Bird}}, \bibinfo {author} {\bibfnamefont {E.}~\bibnamefont {Smith}},\ and\ \bibinfo {author} {\bibfnamefont {Z.}~\bibnamefont {Liang}},\ }\bibfield  {title} {\bibinfo {title} {Coalescence characteristics of bulk nanobubbles in water: A molecular dynamics study coupled with theoretical analysis},\ }\href@noop {} {\bibfield  {journal} {\bibinfo  {journal} {Phys. Rev. Fluids}\ }\textbf {\bibinfo {volume} {6}},\ \bibinfo {pages} {093604} (\bibinfo {year} {2021})}\BibitemShut {NoStop}%
\bibitem [{\citenamefont {Eggers}\ \emph {et~al.}(2024)\citenamefont {Eggers}, \citenamefont {Sprittles},\ and\ \citenamefont {Snoeijer}}]{eggers2024coalescence}%
  \BibitemOpen
  \bibfield  {author} {\bibinfo {author} {\bibfnamefont {J.}~\bibnamefont {Eggers}}, \bibinfo {author} {\bibfnamefont {J.~E.}\ \bibnamefont {Sprittles}},\ and\ \bibinfo {author} {\bibfnamefont {J.~H.}\ \bibnamefont {Snoeijer}},\ }\bibfield  {title} {\bibinfo {title} {Coalescence dynamics},\ }\href@noop {} {\bibfield  {journal} {\bibinfo  {journal} {Annu. Rev. Fluid Mech.}\ }\textbf {\bibinfo {volume} {57}} (\bibinfo {year} {2024})}\BibitemShut {NoStop}%
\bibitem [{\citenamefont {Lohse}\ and\ \citenamefont {Zhang}(2015)}]{lohse2015surface}%
  \BibitemOpen
  \bibfield  {author} {\bibinfo {author} {\bibfnamefont {D.}~\bibnamefont {Lohse}}\ and\ \bibinfo {author} {\bibfnamefont {X.}~\bibnamefont {Zhang}},\ }\bibfield  {title} {\bibinfo {title} {Surface nanobubbles and nanodroplets},\ }\href@noop {} {\bibfield  {journal} {\bibinfo  {journal} {Rev. Mod. Phys.}\ }\textbf {\bibinfo {volume} {87}},\ \bibinfo {pages} {981} (\bibinfo {year} {2015})}\BibitemShut {NoStop}%
\bibitem [{\citenamefont {Chen}\ \emph {et~al.}(2014)\citenamefont {Chen}, \citenamefont {Luo}, \citenamefont {Faraji}, \citenamefont {Feldberg},\ and\ \citenamefont {White}}]{chen2014electrochemical}%
  \BibitemOpen
  \bibfield  {author} {\bibinfo {author} {\bibfnamefont {Q.}~\bibnamefont {Chen}}, \bibinfo {author} {\bibfnamefont {L.}~\bibnamefont {Luo}}, \bibinfo {author} {\bibfnamefont {H.}~\bibnamefont {Faraji}}, \bibinfo {author} {\bibfnamefont {S.~W.}\ \bibnamefont {Feldberg}},\ and\ \bibinfo {author} {\bibfnamefont {H.~S.}\ \bibnamefont {White}},\ }\bibfield  {title} {\bibinfo {title} {Electrochemical measurements of single {H\textsubscript{2}} nanobubble nucleation and stability at {Pt} nanoelectrodes},\ }\href@noop {} {\bibfield  {journal} {\bibinfo  {journal} {J. Phys. Chem. Lett.}\ }\textbf {\bibinfo {volume} {5}},\ \bibinfo {pages} {3539} (\bibinfo {year} {2014})}\BibitemShut {NoStop}%
\bibitem [{\citenamefont {Plimpton}(1995)}]{plimpton1995fast}%
  \BibitemOpen
  \bibfield  {author} {\bibinfo {author} {\bibfnamefont {S.}~\bibnamefont {Plimpton}},\ }\bibfield  {title} {\bibinfo {title} {Fast parallel algorithms for short-range molecular dynamics},\ }\href@noop {} {\bibfield  {journal} {\bibinfo  {journal} {J. Comput. Phys.}\ }\textbf {\bibinfo {volume} {117}},\ \bibinfo {pages} {1} (\bibinfo {year} {1995})}\BibitemShut {NoStop}%
\bibitem [{\citenamefont {Molinero}\ and\ \citenamefont {Moore}(2009)}]{molinero2009water}%
  \BibitemOpen
  \bibfield  {author} {\bibinfo {author} {\bibfnamefont {V.}~\bibnamefont {Molinero}}\ and\ \bibinfo {author} {\bibfnamefont {E.~B.}\ \bibnamefont {Moore}},\ }\bibfield  {title} {\bibinfo {title} {Water modeled as an intermediate element between carbon and silicon},\ }\href@noop {} {\bibfield  {journal} {\bibinfo  {journal} {J. Phys. Chem. B}\ }\textbf {\bibinfo {volume} {113}},\ \bibinfo {pages} {4008} (\bibinfo {year} {2009})}\BibitemShut {NoStop}%
\bibitem [{\citenamefont {Zhang}\ and\ \citenamefont {Lohse}(2025)}]{zhang2025motion}%
  \BibitemOpen
  \bibfield  {author} {\bibinfo {author} {\bibfnamefont {Y.}~\bibnamefont {Zhang}}\ and\ \bibinfo {author} {\bibfnamefont {D.}~\bibnamefont {Lohse}},\ }\bibfield  {title} {\bibinfo {title} {Motion of bulk nanobubbles driven by thermal {Marangoni} flow},\ }\href@noop {} {\bibfield  {journal} {\bibinfo  {journal} {J. Fluid Mech.}\ }\textbf {\bibinfo {volume} {1008}},\ \bibinfo {pages} {A39} (\bibinfo {year} {2025})}\BibitemShut {NoStop}%
\bibitem [{\citenamefont {Shekhar}\ \emph {et~al.}(2013)\citenamefont {Shekhar}, \citenamefont {Nomura}, \citenamefont {Kalia}, \citenamefont {Nakano},\ and\ \citenamefont {Vashishta}}]{shekhar2013nanobubble}%
  \BibitemOpen
  \bibfield  {author} {\bibinfo {author} {\bibfnamefont {A.}~\bibnamefont {Shekhar}}, \bibinfo {author} {\bibfnamefont {K.-i.}\ \bibnamefont {Nomura}}, \bibinfo {author} {\bibfnamefont {R.~K.}\ \bibnamefont {Kalia}}, \bibinfo {author} {\bibfnamefont {A.}~\bibnamefont {Nakano}},\ and\ \bibinfo {author} {\bibfnamefont {P.}~\bibnamefont {Vashishta}},\ }\bibfield  {title} {\bibinfo {title} {Nanobubble collapse on a silica surface in water: Billion-atom reactive molecular dynamics simulations},\ }\href@noop {} {\bibfield  {journal} {\bibinfo  {journal} {Phys. Rev. Lett.}\ }\textbf {\bibinfo {volume} {111}},\ \bibinfo {pages} {184503} (\bibinfo {year} {2013})}\BibitemShut {NoStop}%
\bibitem [{\citenamefont {Giacomello}\ \emph {et~al.}(2012)\citenamefont {Giacomello}, \citenamefont {Chinappi}, \citenamefont {Meloni},\ and\ \citenamefont {Casciola}}]{giacomello2012metastable}%
  \BibitemOpen
  \bibfield  {author} {\bibinfo {author} {\bibfnamefont {A.}~\bibnamefont {Giacomello}}, \bibinfo {author} {\bibfnamefont {M.}~\bibnamefont {Chinappi}}, \bibinfo {author} {\bibfnamefont {S.}~\bibnamefont {Meloni}},\ and\ \bibinfo {author} {\bibfnamefont {C.~M.}\ \bibnamefont {Casciola}},\ }\bibfield  {title} {\bibinfo {title} {Metastable wetting on superhydrophobic surfaces: Continuum and atomistic views of the {Cassie-Baxter-Wenzel} transition},\ }\href@noop {} {\bibfield  {journal} {\bibinfo  {journal} {Phys. Rev. Lett.}\ }\textbf {\bibinfo {volume} {109}},\ \bibinfo {pages} {226102} (\bibinfo {year} {2012})}\BibitemShut {NoStop}%
\bibitem [{\citenamefont {Xiang}\ \emph {et~al.}(2017)\citenamefont {Xiang}, \citenamefont {Huang}, \citenamefont {Lv}, \citenamefont {Xue}, \citenamefont {Su},\ and\ \citenamefont {Duan}}]{xiang2017ultimate}%
  \BibitemOpen
  \bibfield  {author} {\bibinfo {author} {\bibfnamefont {Y.}~\bibnamefont {Xiang}}, \bibinfo {author} {\bibfnamefont {S.}~\bibnamefont {Huang}}, \bibinfo {author} {\bibfnamefont {P.}~\bibnamefont {Lv}}, \bibinfo {author} {\bibfnamefont {Y.}~\bibnamefont {Xue}}, \bibinfo {author} {\bibfnamefont {Q.}~\bibnamefont {Su}},\ and\ \bibinfo {author} {\bibfnamefont {H.}~\bibnamefont {Duan}},\ }\bibfield  {title} {\bibinfo {title} {Ultimate stable underwater superhydrophobic state},\ }\href@noop {} {\bibfield  {journal} {\bibinfo  {journal} {Phys. Rev. Lett.}\ }\textbf {\bibinfo {volume} {119}},\ \bibinfo {pages} {134501} (\bibinfo {year} {2017})}\BibitemShut {NoStop}%
\bibitem [{\citenamefont {Oguz}\ and\ \citenamefont {Prosperetti}(1993)}]{oguz1993dynamics}%
  \BibitemOpen
  \bibfield  {author} {\bibinfo {author} {\bibfnamefont {H.~N.}\ \bibnamefont {Oguz}}\ and\ \bibinfo {author} {\bibfnamefont {A.}~\bibnamefont {Prosperetti}},\ }\bibfield  {title} {\bibinfo {title} {Dynamics of bubble growth and detachment from a needle},\ }\href@noop {} {\bibfield  {journal} {\bibinfo  {journal} {J. Fluid Mech.}\ }\textbf {\bibinfo {volume} {257}},\ \bibinfo {pages} {111} (\bibinfo {year} {1993})}\BibitemShut {NoStop}%
\bibitem [{\citenamefont {Brenner}\ \emph {et~al.}(2002)\citenamefont {Brenner}, \citenamefont {Hilgenfeldt},\ and\ \citenamefont {Lohse}}]{brenner2002single}%
  \BibitemOpen
  \bibfield  {author} {\bibinfo {author} {\bibfnamefont {M.~P.}\ \bibnamefont {Brenner}}, \bibinfo {author} {\bibfnamefont {S.}~\bibnamefont {Hilgenfeldt}},\ and\ \bibinfo {author} {\bibfnamefont {D.}~\bibnamefont {Lohse}},\ }\bibfield  {title} {\bibinfo {title} {Single-bubble sonoluminescence},\ }\href@noop {} {\bibfield  {journal} {\bibinfo  {journal} {Rev. Mod. Phys.}\ }\textbf {\bibinfo {volume} {74}},\ \bibinfo {pages} {425} (\bibinfo {year} {2002})}\BibitemShut {NoStop}%
\bibitem [{\citenamefont {Feng}\ and\ \citenamefont {Leal}(1997)}]{feng1997nonlinear}%
  \BibitemOpen
  \bibfield  {author} {\bibinfo {author} {\bibfnamefont {Z.}~\bibnamefont {Feng}}\ and\ \bibinfo {author} {\bibfnamefont {L.}~\bibnamefont {Leal}},\ }\bibfield  {title} {\bibinfo {title} {Nonlinear bubble dynamics},\ }\href@noop {} {\bibfield  {journal} {\bibinfo  {journal} {Annu. Rev. Fluid Mech.}\ }\textbf {\bibinfo {volume} {29}},\ \bibinfo {pages} {201} (\bibinfo {year} {1997})}\BibitemShut {NoStop}%
\bibitem [{\citenamefont {Plesset}\ and\ \citenamefont {Prosperetti}(1977)}]{plesset1977bubble}%
  \BibitemOpen
  \bibfield  {author} {\bibinfo {author} {\bibfnamefont {M.~S.}\ \bibnamefont {Plesset}}\ and\ \bibinfo {author} {\bibfnamefont {A.}~\bibnamefont {Prosperetti}},\ }\bibfield  {title} {\bibinfo {title} {Bubble dynamics and cavitation},\ }\href@noop {} {\bibfield  {journal} {\bibinfo  {journal} {Annu. Rev. Fluid Mech.}\ }\textbf {\bibinfo {volume} {9}},\ \bibinfo {pages} {145} (\bibinfo {year} {1977})}\BibitemShut {NoStop}%
\bibitem [{\citenamefont {Prosperetti}\ \emph {et~al.}(1988)\citenamefont {Prosperetti}, \citenamefont {Crum},\ and\ \citenamefont {Commander}}]{prosperetti1988nonlinear}%
  \BibitemOpen
  \bibfield  {author} {\bibinfo {author} {\bibfnamefont {A.}~\bibnamefont {Prosperetti}}, \bibinfo {author} {\bibfnamefont {L.~A.}\ \bibnamefont {Crum}},\ and\ \bibinfo {author} {\bibfnamefont {K.~W.}\ \bibnamefont {Commander}},\ }\bibfield  {title} {\bibinfo {title} {Nonlinear bubble dynamics},\ }\href@noop {} {\bibfield  {journal} {\bibinfo  {journal} {J. Acoust. Soc. Am.}\ }\textbf {\bibinfo {volume} {83}},\ \bibinfo {pages} {502} (\bibinfo {year} {1988})}\BibitemShut {NoStop}%
\bibitem [{\citenamefont {Cheng}\ \emph {et~al.}(2016)\citenamefont {Cheng}, \citenamefont {Xu},\ and\ \citenamefont {Sui}}]{cheng2016numerical}%
  \BibitemOpen
  \bibfield  {author} {\bibinfo {author} {\bibfnamefont {Y.}~\bibnamefont {Cheng}}, \bibinfo {author} {\bibfnamefont {J.}~\bibnamefont {Xu}},\ and\ \bibinfo {author} {\bibfnamefont {Y.}~\bibnamefont {Sui}},\ }\bibfield  {title} {\bibinfo {title} {Numerical investigation of coalescence-induced droplet jumping on superhydrophobic surfaces for efficient dropwise condensation heat transfer},\ }\href@noop {} {\bibfield  {journal} {\bibinfo  {journal} {Int. J. Heat Mass Transf.}\ }\textbf {\bibinfo {volume} {95}},\ \bibinfo {pages} {506} (\bibinfo {year} {2016})}\BibitemShut {NoStop}%
\bibitem [{\citenamefont {Perumanath}\ \emph {et~al.}(2020)\citenamefont {Perumanath}, \citenamefont {Borg}, \citenamefont {Sprittles},\ and\ \citenamefont {Enright}}]{perumanath2020molecular}%
  \BibitemOpen
  \bibfield  {author} {\bibinfo {author} {\bibfnamefont {S.}~\bibnamefont {Perumanath}}, \bibinfo {author} {\bibfnamefont {M.~K.}\ \bibnamefont {Borg}}, \bibinfo {author} {\bibfnamefont {J.~E.}\ \bibnamefont {Sprittles}},\ and\ \bibinfo {author} {\bibfnamefont {R.}~\bibnamefont {Enright}},\ }\bibfield  {title} {\bibinfo {title} {Molecular physics of jumping nanodroplets},\ }\href@noop {} {\bibfield  {journal} {\bibinfo  {journal} {Nanoscale}\ }\textbf {\bibinfo {volume} {12}},\ \bibinfo {pages} {20631} (\bibinfo {year} {2020})}\BibitemShut {NoStop}%
\bibitem [{\citenamefont {Huang}\ \emph {et~al.}(2019)\citenamefont {Huang}, \citenamefont {Huang},\ and\ \citenamefont {Xu}}]{huang2019energy}%
  \BibitemOpen
  \bibfield  {author} {\bibinfo {author} {\bibfnamefont {J.-J.}\ \bibnamefont {Huang}}, \bibinfo {author} {\bibfnamefont {H.}~\bibnamefont {Huang}},\ and\ \bibinfo {author} {\bibfnamefont {J.-J.}\ \bibnamefont {Xu}},\ }\bibfield  {title} {\bibinfo {title} {Energy-based modeling of micro-and nano-droplet jumping upon coalescence on superhydrophobic surfaces},\ }\href@noop {} {\bibfield  {journal} {\bibinfo  {journal} {Appl. Phys. Lett.}\ }\textbf {\bibinfo {volume} {115}} (\bibinfo {year} {2019})}\BibitemShut {NoStop}%
\bibitem [{\citenamefont {Brennen}(2014)}]{brennen2014cavitation}%
  \BibitemOpen
  \bibfield  {author} {\bibinfo {author} {\bibfnamefont {C.~E.}\ \bibnamefont {Brennen}},\ }\href@noop {} {\emph {\bibinfo {title} {Cavitation and bubble dynamics}}}\ (\bibinfo  {publisher} {Cambridge University Press},\ \bibinfo {year} {2014})\BibitemShut {NoStop}%
\bibitem [{\citenamefont {Fuster}\ and\ \citenamefont {Popinet}(2018)}]{fuster2018all}%
  \BibitemOpen
  \bibfield  {author} {\bibinfo {author} {\bibfnamefont {D.}~\bibnamefont {Fuster}}\ and\ \bibinfo {author} {\bibfnamefont {S.}~\bibnamefont {Popinet}},\ }\bibfield  {title} {\bibinfo {title} {An all-{Mach} method for the simulation of bubble dynamics problems in the presence of surface tension},\ }\href@noop {} {\bibfield  {journal} {\bibinfo  {journal} {J. Comput. Phys.}\ }\textbf {\bibinfo {volume} {374}},\ \bibinfo {pages} {752} (\bibinfo {year} {2018})}\BibitemShut {NoStop}%
\bibitem [{\citenamefont {Popinet}(2018)}]{popinet2018numerical}%
  \BibitemOpen
  \bibfield  {author} {\bibinfo {author} {\bibfnamefont {S.}~\bibnamefont {Popinet}},\ }\bibfield  {title} {\bibinfo {title} {Numerical models of surface tension},\ }\href@noop {} {\bibfield  {journal} {\bibinfo  {journal} {Annu. Rev. Fluid Mech.}\ }\textbf {\bibinfo {volume} {50}},\ \bibinfo {pages} {49} (\bibinfo {year} {2018})}\BibitemShut {NoStop}%
\bibitem [{\citenamefont {Nag}\ \emph {et~al.}(2021)\citenamefont {Nag}, \citenamefont {Tomo}, \citenamefont {Teshima}, \citenamefont {Takahashi},\ and\ \citenamefont {Kohno}}]{nag2021dynamic}%
  \BibitemOpen
  \bibfield  {author} {\bibinfo {author} {\bibfnamefont {S.}~\bibnamefont {Nag}}, \bibinfo {author} {\bibfnamefont {Y.}~\bibnamefont {Tomo}}, \bibinfo {author} {\bibfnamefont {H.}~\bibnamefont {Teshima}}, \bibinfo {author} {\bibfnamefont {K.}~\bibnamefont {Takahashi}},\ and\ \bibinfo {author} {\bibfnamefont {M.}~\bibnamefont {Kohno}},\ }\bibfield  {title} {\bibinfo {title} {Dynamic interplay between interfacial nanobubbles: Oversaturation promotes anisotropic depinning and bubble coalescence},\ }\href@noop {} {\bibfield  {journal} {\bibinfo  {journal} {Phys. Chem. Chem. Phys.}\ }\textbf {\bibinfo {volume} {23}},\ \bibinfo {pages} {24652} (\bibinfo {year} {2021})}\BibitemShut {NoStop}%
\bibitem [{\citenamefont {Shin}\ \emph {et~al.}(2015)\citenamefont {Shin}, \citenamefont {Park}, \citenamefont {Kim}, \citenamefont {Kim}, \citenamefont {Kang}, \citenamefont {Lee}, \citenamefont {Cho}, \citenamefont {Hong},\ and\ \citenamefont {Novoselov}}]{shin2015growth}%
  \BibitemOpen
  \bibfield  {author} {\bibinfo {author} {\bibfnamefont {D.}~\bibnamefont {Shin}}, \bibinfo {author} {\bibfnamefont {J.~B.}\ \bibnamefont {Park}}, \bibinfo {author} {\bibfnamefont {Y.-J.}\ \bibnamefont {Kim}}, \bibinfo {author} {\bibfnamefont {S.~J.}\ \bibnamefont {Kim}}, \bibinfo {author} {\bibfnamefont {J.~H.}\ \bibnamefont {Kang}}, \bibinfo {author} {\bibfnamefont {B.}~\bibnamefont {Lee}}, \bibinfo {author} {\bibfnamefont {S.-P.}\ \bibnamefont {Cho}}, \bibinfo {author} {\bibfnamefont {B.~H.}\ \bibnamefont {Hong}},\ and\ \bibinfo {author} {\bibfnamefont {K.~S.}\ \bibnamefont {Novoselov}},\ }\bibfield  {title} {\bibinfo {title} {Growth dynamics and gas transport mechanism of nanobubbles in graphene liquid cells},\ }\href@noop {} {\bibfield  {journal} {\bibinfo  {journal} {Nat. Commun.}\ }\textbf {\bibinfo {volume} {6}},\ \bibinfo {pages} {6068} (\bibinfo {year} {2015})}\BibitemShut {NoStop}%
\bibitem [{\citenamefont {Chan}\ \emph {et~al.}(2011)\citenamefont {Chan}, \citenamefont {Klaseboer},\ and\ \citenamefont {Manica}}]{C0SM00812E}%
  \BibitemOpen
  \bibfield  {author} {\bibinfo {author} {\bibfnamefont {D.~Y.~C.}\ \bibnamefont {Chan}}, \bibinfo {author} {\bibfnamefont {E.}~\bibnamefont {Klaseboer}},\ and\ \bibinfo {author} {\bibfnamefont {R.}~\bibnamefont {Manica}},\ }\bibfield  {title} {\bibinfo {title} {Film drainage and coalescence between deformable drops and bubbles},\ }\href@noop {} {\bibfield  {journal} {\bibinfo  {journal} {Soft Matter}\ }\textbf {\bibinfo {volume} {7}},\ \bibinfo {pages} {2235} (\bibinfo {year} {2011})}\BibitemShut {NoStop}%
\bibitem [{\citenamefont {Forel}\ \emph {et~al.}(2019)\citenamefont {Forel}, \citenamefont {Dollet}, \citenamefont {Langevin},\ and\ \citenamefont {Rio}}]{PhysRevLett.122.088002}%
  \BibitemOpen
  \bibfield  {author} {\bibinfo {author} {\bibfnamefont {E.}~\bibnamefont {Forel}}, \bibinfo {author} {\bibfnamefont {B.}~\bibnamefont {Dollet}}, \bibinfo {author} {\bibfnamefont {D.}~\bibnamefont {Langevin}},\ and\ \bibinfo {author} {\bibfnamefont {E.}~\bibnamefont {Rio}},\ }\bibfield  {title} {\bibinfo {title} {Coalescence in two-dimensional foams: A purely statistical process dependent on film area},\ }\href@noop {} {\bibfield  {journal} {\bibinfo  {journal} {Phys. Rev. Lett.}\ }\textbf {\bibinfo {volume} {122}},\ \bibinfo {pages} {088002} (\bibinfo {year} {2019})}\BibitemShut {NoStop}%
\end{thebibliography}
\end{document}